\documentclass[preprint]{aastex}

\shortauthors{Manset and Bastien}
\shorttitle{Polarimetric observations of MWC~1080}

\begin{document}

%%%%%%%%%%%%%%%%%%%%%%%%%%%%%%%%%%%%%%%%%%%%%%%%%%%%%%%%%%%%%%%%%%%%%%%%
% TITLE PAGE
%%%%%%%%%%%%%%%%%%%%%%%%%%%%%%%%%%%%%%%%%%%%%%%%%%%%%%%%%%%%%%%%%%%%%%%%
\title{Polarimetric variations of binary stars. III Periodic
polarimetric variations of the Herbig Ae/Be star MWC~1080\footnote{Based
(in part) on observations collected with the 2m Bernard-Lyot telescope
(TBL) operated by INSU/CNRS and Pic-du-Midi Observatory (CNRS USR
5026).}}
\author{N. Manset\altaffilmark{2} and P. Bastien} \affil{Universit\'e de
Montr\'eal, D\'epartement de Physique, and Observatoire du Mont
M\'egantic, C.P. 6128, Succ. Centre-ville, Montr\'eal, QC, H3C 3J7,
Canada} \email{manset@cfht.hawaii.edu,
bastien@astro.umontreal.ca}
\altaffiltext{2}{Now at: Canada-France-Hawaii Telescope Corporation,
P.O. Box 1597, Kamuela, HI 96743, USA} 

%%%%%%%%%%%%%%%%%%%%%%%%%%%%%%%%%%%%%%%%%%%%%%%%%%%%%%%%%%%%%%%%%%%%%%%%
% ABSTRACT and SUBJECT HEADINGS
%%%%%%%%%%%%%%%%%%%%%%%%%%%%%%%%%%%%%%%%%%%%%%%%%%%%%%%%%%%%%%%%%%%%%%%%
\begin{abstract}
We present polarimetric observations of a massive pre-main sequence
short-period binary star of the Herbig~Ae/Be type, MWC~1080. The mean
polarization at $7660\;$\AA\ is 1.60\% at 81.6\arcdeg, or 0.6\% at
139\arcdeg\ if an estimate of the interstellar polarization is
subtracted. The intrinsic polarization points to an
asymmetric geometry of the circumstellar or circumbinary environment
while the 139\arcdeg\ intrinsic position angle traces the axis of
symmetry of the system and is perpendicular to the position angle of the
outflow cavity. The polarization and its position angle are clearly
variable, at all wavelengths, and on time scales of hours, days, months,
and years. Stochastic variability is accompanied by periodic variations
caused by the orbital motion of the stars in their dusty
environment. These periodic polarimetric variations are the first
phased-locked ones detected for a pre-main sequence binary. The
variations are not simply double-periodic (seen twice per orbit) but
include single-periodic (seen once per orbit) and higher-order
variations. The presence of single-periodic variations could be due to
non equal mass stars, the presence of dust grains, an asymmetric
configuration of the circumstellar or circumbinary material, or the
eccentricity of the orbit.  MWC~1080 is an eclipsing binary with primary
and secondary eclipses occurring at phases 0.0 and 0.55. The signatures
of the eclipses are seen in the polarimetric observations.
\end{abstract}


%%%%%%%%%%%%%%%%%%%%%%%%%%%%%%%%%%%%%%%%%%%%%%%%%%%%%%%%%%%%%%%%%%%%%%%%
% INTRODUCTION 
%%%%%%%%%%%%%%%%%%%%%%%%%%%%%%%%%%%%%%%%%%%%%%%%%%%%%%%%%%%%%%%%%%%%%%%%
\section{Introduction}

Pre-main sequence (PMS) stars are objects still contracting to the main
sequence and generally surrounded by disks and/or envelopes of
circumstellar dust and gas, which represent remnant material from their
still-ongoing formation. This circumstellar matter is responsible for
phenomenons such as emission lines, P~Cygni profiles, IR and UV
excesses, and polarization. This polarization is produced by scattering
on dust grains and has been known for a number of years (see Bastien
1996 for a review).

It is also well known that when a binary star (of any evolutionary
status) is surrounded by circumstellar matter, its polarization varies
as a function of the orbital period. Some models seek to reproduce these
variations (see for example Rudy \& Kemp 1978; Brown, McLean, \& Emslie
1978) in order to find the orbital inclination of these binary
systems. In particular, the work from Brown et al.  (1978) (hereafter
referred to as BME) uses first and second-order Fourier analysis of the
Stokes curves to give, in addition to the orbital inclination, moments
related to the distribution of the scatterers in the circumstellar and
circumbinary environment.

It could be very interesting to use the BME formalism to find the orbital
inclination of PMS binaries, since spectroscopic observations coupled
with an orbital inclination can yield the absolute mass of each
star. These masses could then be compared with theoretical models of the
formation of binary stars, and to masses derived from other types of
observations (photometry and theoretical evolutionary tracks for
example).

However, the BME formalism was developed for Thomson scattering in optically
thin envelopes, and for binaries in circular orbits. Since polarization
in PMS stars is produced by scattering on dust grains, and most of the
known PMS spectroscopic binaries have eccentric orbits, the BME
formalism can not be used a priori. Studies were undertaken to verify
the applicability of the BME formalism for Mie scattering and eccentric
orbits (see Manset \& Bastien 2000, 2001, hereafter referred to as Paper
I and II respectively) and have shown that the BME analysis can still be
applied in those cases, with a few limitations.

In this context, we have obtained polarimetric observations of 24 PMS
spectroscopic binaries, following (with $\gtrsim 10$ observations) the
shortest-period ($P \lesssim\; 35\;$d) and brightest ($V \lesssim 12.0$)
ones. The observations were used to study the general polarimetric
variability of these systems, correlations between these variations and
physical characteristics, periodicity in the variations, and to try to
determine the orbital inclination.

In this paper, we present the observational method, method of
determination of the interstellar polarization, analysis of the
polarimetric variability and of the periodic variations, along with data
for the only Herbig Ae/Be star of our sample, MWC~1080. The observations
and discussion for the rest of the sample will be presented in future
papers.
  
%%%%%%%%%%%%%%%%%%%%%%%%%%%%%%%%%%%%%%%%%%%%%%%%%%%%%%%%%%%%%%%%%%%%%%%%
% OBSERVATIONS
%%%%%%%%%%%%%%%%%%%%%%%%%%%%%%%%%%%%%%%%%%%%%%%%%%%%%%%%%%%%%%%%%%%%%%%%
\section{Observations}
Polarimetric data were taken at the Observatoire du Mont M\'egantic
(OMM), Canada, with Beauty and The Beast, a two-channel photo-electric
polarimeter used at the f/8 Cassegrain focus of the 1.6m telescope. In this
polarimeter, the analyzer is a Wollaston prism which is used in
combination with a Pockels cell acting as a variable quarter-wave plate,
and an additional quarter-wave plate for linear polarization
measurements. See Manset \& Bastien (1995) and Manset (2000) for details
about the instrument. The high voltage supplied to the Pockels cell is
switched at 62.5~Hz to beat down the effects of variable seeing and
transparency (Serkowski 1974).

The data were calibrated for instrumental efficiency, instrumental
polarization (due to the telescope's mirrors), and origin of position
angles using, respectively, a Glan-Thomson polarizing prism, non-polarized
and polarized standard stars. Calibration measurements taken between
1995 and 1999 reveal a very low and stable instrumental polarization
under 0.025\% $\pm\;0.010$\%, and a relatively stable correction for the
position angle between $-37\arcdeg$ and $-33\arcdeg$ from one observing
run to the other.

The observational errors are calculated from photon statistics, but
include uncertainties introduced by the polarimetric efficiency of the
instrument, the instrumental polarization, and the calibration for the
origin of the position angle.  The relative errors in position angle can
be as low as 0.1\arcdeg, but due to instrumental effects, systematic
errors, and the calibration procedure itself, the absolute errors on the
position angles are of the order of 1$^\circ$. Details on the data
reduction, stability of the instrument, and calibrations are given by
Manset (2000).

We have also obtained data using the polarimeter STERENN at the 2-meter
Bernard-Lyot Telescope of the Observatoire du Pic-du-Midi (OPdM),
France. This 2-channel polarimeter uses a half-wave plate rotating at
20~Hz and a Wollaston prism, along with red photo-diodes. Instrumental
efficiency is measured with polarized standard stars, and the
instrumental polarization, with non-polarized standard stars. The
observational errors are calculated from the fit of the observations
taken at a frequency of 20~Hz to a sinusoid function.

%%%%%%%%%%%%%%%%%%%%%%%%%%%%%%%%%%%%%%%%%%%%%%%%%%%%%%%%%%%%%%%%%%%%%%%%
%%% IS POLARIZATION
%%%%%%%%%%%%%%%%%%%%%%%%%%%%%%%%%%%%%%%%%%%%%%%%%%%%%%%%%%%%%%%%%%%%%%%%
\section{Estimation of the interstellar polarization}
Polarimetric observations are in general a sum of intrinsic and
interstellar polarizations. It is useful to estimate the importance of
interstellar polarization in each measurement made.  One can start by
looking at the Mathewson et al.  (1978) catalog, which contains over 7000
polarimetric observations of bright early-type stars, and compare the
level of polarization and position angle of neighboring stars
\footnote{One could also use the more recent catalog by Heiles (2000)
which contains observations for over 9000 stars.}. However, this can be
misleading since interstellar polarization depends on the distance; a
target with high polarization could simply be further away and not
necessarily intrinsically polarized.

Therefore, for each observed PMS binary, the catalog was scanned to
select at least 20 close stars {\it with a similar distance modulus};
depending on the stellar density and number of measurements in the
catalog, this led to the selection of a region between 1 and 15\arcdeg\
in radius around the target.

The stars selected from the catalog are used to find the ratio
$P/E(B-V)$, and finally, based on an extinction value for our target and
assuming this extinction is of interstellar origin only, an estimate of
the interstellar polarization for the target is calculated, along with
the average interstellar polarization angle. This angle is calculated
with a simple average and with a distance-weighted average of the
polarization angles of all the stars selected. If the alignment is good
over all of the region studied, the two values will be similar to within
$\pm\;10\arcdeg$; if they are very different, it means the alignment is
not very good and it is harder to find an average interstellar
polarization angle.

If the position angles for the interstellar polarization and for the
target are different, it points to an intrinsic origin for at least part
of the polarization measured. Intrinsic polarization is also deduced
from polarimetric variability. Histograms and maps of the polarization
and its position angle are also used to estimate the importance of
interstellar polarization.

%%%%%%%%%%%%%%%%%%%%%%%%%%%%%%%%%%%%%%%%%%%%%%%%%%%%%%%%%%%%%%%%%%%%%%%%
%%% POLARIMETRIC VARIABILITY
%%%%%%%%%%%%%%%%%%%%%%%%%%%%%%%%%%%%%%%%%%%%%%%%%%%%%%%%%%%%%%%%%%%%%%%%
\section{Variability tests \label{section-var-tests}}

Since a majority of single PMS stars are variable polarimetrically
(Bastien 1982; Drissen, Bastien, \& St-Louis 1989; M\'enard \& Bastien
1992), we expected PMS binaries, and MWC~1080 in particular, to also be
polarimetrically variable, either periodically or not. Various
statistical tests were applied to check the polarimetric variability or
stability of PMS binaries; the details of those variability tests and
the results for all the binaries of our sample will be presented in
future papers.

Many PMS binary stars show odd observations with polarization levels
and/or position angles well below or above the bulk of the data, and
MWC~1080 is no exception. These data were removed before testing for
variability. We believe these observations were due to some
eruption-like events or significant modifications in the circumstellar
environment (formation/destruction of condensations, accretion events),
and not because of an instrumental problem: a close examination of
polarization observations taken over 5 years of non-polarized standard
stars (84 observations), polarized standard stars (53 observations), and
3 stars that were followed for many consecutive hours (121 observations)
did not show odd observations like the ones we repeatedly saw for PMS
binaries in general, and MWC~1080 in particular (Manset 2000).

%%%%%%%%%%%%%%%%%%%%%%%%%%%%%%%%%%%%%%%%%%%%%%%%%%%%%%%%%%%%%%%%%%%%%%%%
%%% PERIODIC VARIATIONS
%%%%%%%%%%%%%%%%%%%%%%%%%%%%%%%%%%%%%%%%%%%%%%%%%%%%%%%%%%%%%%%%%%%%%%%%
\section{Periodic variations \label{p3-sec-pervar}}
In addition to stochastic polarimetric variability, which is a general
property of single PMS stars, PMS binaries will also present periodic
polarimetric variations caused by the orbital motion, even if in some
cases the amplitude may be too small to be detected with currently
available instruments or masked by non-periodic or pseudo-periodic
variations. In the case of Mie scattering, we have also shown (Paper II)
that dust grains, which are the main producers of the polarization, are
less efficient polarizers and produce smaller amplitude variations than
electrons; this is an indication that periodic polarimetric variations
could be more difficult to observe in PMS binaries than in, for example,
hot stars surrounded by electrons, which can easily show variations of a
few tenths of a percent (see for example Robert et al.  1990; Robert et
al.  1992).

Using the known orbital period, we calculate the orbital phase for each
observation of each star and plot $P$, $\theta$ and the Stokes
parameters $Q$ and $U$ as a function of the orbital phase. When enough
data are available, observations are represented as first and second
harmonics of $\lambda=2\pi\phi$, where $\phi$ is the orbital phase, $0 <
\phi < 1$:\\ 
\begin{eqnarray}
Q &=& q_0 + q_1 \cos \lambda + q_2 \sin \lambda + q_3 \cos 2\lambda +
q_4 \sin 2\lambda, \label{p3-eq-qfit}\\
U &=& u_0 + u_1 \cos \lambda + u_2 \sin \lambda + u_3 \cos 2\lambda +
u_4 \sin 2\lambda. \label{p3-eq-ufit}
\end{eqnarray}
The coefficients of this fit are then used to find the orbital
inclination, using the first or second-order Fourier coefficients,
although it is usually expected that second order variations will
dominate (BME):\\
\begin{eqnarray}
\left[ \frac{1-\cos i}{1+\cos i} \right]^2 &=& \frac{(u_1+q_2)^2 +
(u_2-q_1)^2}{(u_2+q_1)^2 + (u_1-q_2)^2} \label{EQ-iO1-p3},\\
\left[ \frac{1-\cos i}{1+\cos i} \right]^4 &=& \frac{(u_3+q_4)^2 +
(u_4-q_3)^2}{(u_4+q_3)^2 + (u_3-q_4)^2} \label{EQ-iO2-p3}.
\end{eqnarray}

Non-periodic phenomenons such as eruptive events, transient stellar
spots, variable accretion, and rearrangements of the circumstellar or
circumbinary material can cause pseudo-periodic polarimetric variations
that may mask the strictly periodic ones, especially if the observations
are taken over many orbital periods, as is the case here.

When periodic variations are detected, we use a Phase Dispersion Method
(Stellingwerf 1978) and a Lomb normalized periodogram algorithm (Press
et al.  1997) to investigate the significance of this periodicity. The
Phase Dispersion Method (PDM) is a least-squared fitting technique
suited for non-sinusoidal time variations covered by irregularly spaced
observations, and finds the period that produces the least scatter about
the mean curve. The Lomb normalized periodogram (LNP) method is more
powerful than Fast Fourier Transform methods for uneven sampling, but
still assumes the curve is sinusoidal, which may not be always
appropriate for the polarimetric observations presented here.

%%%%%%%%%%%%%%%%%%%%%%%%%%%%%%%%%%%%%%%%%%%%%%%%%%%%%%%%%%%%%%%%%%%%%%%%
%%%%%%%%%%%%%%%%%%%%%%%%%%%%%%%%%%%%%%%%%%%%%%%%%%%%%%%%%%%%%%%%%%%%%%%%
%%%%%%%%%%%%%%%%%%%%%%%%%%%%%%%%%%%%%%%%%%%%%%%%%%%%%%%%%%%%%%%%%%%%%%%%
% SECTION:           M W C   1 0 8 0 
\section{MWC~1080 = HBC~318 = V628~Cas}

%%%%%%%%%%%%%%%%%%%%%%%%%
\subsection{General characteristics}
Herbig~Ae/Be stars are the higher-mass counterparts ($2\;$M$_{\sun}
\lesssim$ M $\lesssim 10$ M$_{\sun}$) to the T~Tauri stars. These
objects are also associated with nebulosity and surrounded by
circumstellar material; for a review, see Catala (1989). The emission in
IR and mm-wave spectral domains can be explained by the presence of
disks, with possibly central holes, and of envelopes (Miroshnichenko et
al. 1999).

MWC~1080 is a luminous Herbig~Ae/Be object of the Be type with $4\times
10^4$ L$_{\sun}$, and a strong wind (Henning et al. 1998). Its kinematic
distance was determined to be 2.2~kpc (Levreault 1985), and confirmed by
photometry of neighboring stars by Grankin et al. (1992). There is a
third star $0\farcs76$ away, west of MWC~1080 (Leinert et al.  1994),
and a poorly collimated Herbig-Haro flow (Poetzel, Mundt, \& Ray 1992);
the proposed outflow cavity has a position angle of $\approx 60\arcdeg$.

MWC~1080's spectrum is characterized by strong emission lines
(H$\alpha$, H$\beta$, \ion{Fe}{2}), some with P~Cygni profiles, but few
absorption lines (Herbig 1960; Poetzel et al. 1992). The strong emission
lines imply the presence of a large amount of circumstellar gas in the
system.

The emission-line profiles have P Cygni shapes which would indicate that
the gaseous envelope of MWC1080 is rather nearly spherical or at least
not highly flattened, implying a small contribution to the optical
polarization.  Moreover, from spectroscopic data (Corporon \& Lagrange
1999), one could assume that the envelope is not very asymmetric. On the
other hand, according to Hillenbrand et al. (1992), MWC~1080 would
belong to the Group~I Herbig~Ae/Be stars, and thus have a geometrically
flat, optically thick circumstellar accretion disk, with optically thin
inner regions from the stellar surface up to several stellar radii. This
classification could change if other sources (like the other component
of this multiple system) included in the beams or apertures contribute
significantly to the observed emission.

Harvey, Thronson, \& Gatley (1979) have reported an extended
far-infrared emission associated with this object. An $8.8\; \micron$
map by Deutsch et al.  (1995) shows extended elliptical emission
4\arcsec\ or 4000~AU in diameter. At $100\; \micron$, the emission is
also extended (Di Francesco et al.  1994). There is a bipolar molecular
outflow (Yoshida et al.  1991).

MWC~1080 is an eclipsing binary with an orbital period of $2.886926\;$d
(Grankin et al.  1992; Shevchenko et al.  1994). Since the photometric
variations are close to sinusoidal (Grankin et al. 1992), it is thought
that the masses are very similar. The asymmetric light curve and
different height peaks indicate an eccentric orbit with $e \approx
0.2\;$--$\;0.5$, whereas the moderate amplitude is due to a perceptible
inclination of the orbital plane, i.e., the system is not seen exactly
edge-on (Grankin et al. 1992). 

Hillenbrand et al.  (1992) measured a polarization of 1.79\% at
75\arcdeg\ in a $V$ filter, estimated the interstellar polarization from
neighboring stars to be 2.58\% at 71\arcdeg, therefore giving an
intrinsic polarization of 0.79\% \footnote{As given in their paper. See
discussion in Section~\ref{section-is-pol} concerning this
subtraction.}. On the other hand, Garrison \& Anderson (1978), using a
different method, determined a different interstellar polarization of
2.00\% at 168\arcdeg, giving an intrinsic $V$ polarization of 3.84\% at
76\arcdeg.

%%%%%%%%%%%%%%%%%%%%%%%%%%%%%%%%%%%%%%%%%%%%%%%%%%%%%%%%%%%%%%%%%%%%%%%%
%  %%%   %%%   %%%   %%%   %%%   %%%   %%%   %%%   %%%   %%%   %%%   %%%   
%  %%%   %%%   %%%   %%%   %%%   %%%   %%%   %%%   %%%   %%%   %%%   %%%   
% 			MWC1080 - POLARIMETRY
%  %%%   %%%   %%%   %%%   %%%   %%%   %%%   %%%   %%%   %%%   %%%   %%%   
%  %%%   %%%   %%%   %%%   %%%   %%%   %%%   %%%   %%%   %%%   %%%   %%%   
%
\subsection{Polarimetric characteristics}

%%%%%%%%%%%%%%%%%%%%%%%%%
\subsubsection{Polarimetric observations}
MWC~1080 was observed at the OMM (Canada) between August 1995
and June 1999, using a $8\farcs2$ aperture hole and a broad red filter
(filter RG645 centered on $7660\;$\AA, with $2410\;$\AA\ FWHM). These
data are presented in Table~\ref{Tab-data} where we give the Universal
and Julian dates, the orbital phase, polarization and position angle
along with their uncertainties.

The observations thus include the third star, located at $0\farcs76$
from the spectroscopic binary, or $1700\;$AU from the binary, assuming a
distance of $2.2\;$kpc (Levreault 1985; Grankin et al. 1992). Since this
star is located far away, it can contribute to the observations by
adding a constant polarization, but it should not introduce variability
unless it experiences some stochastic phenomenon such as flare-like
eruptions or stellar spots.

Figure~\ref{Fig-MWC1080_all} presents these observations, with the
polarization $P$, position angle $\theta$ and the Stokes parameters $Q$
and $U$ shown as a function of the orbital phase. The solid lines are
the Fourier fits made according to Equations~\ref{p3-eq-qfit} and
\ref{p3-eq-ufit}. As can be seen on that figure, MWC~1080 presents a few
(4) observations which stand out from the rest of the data, and which
were removed for the variability analysis and the fitting procedure
(data from 1996, June 2, July 9, September 1 and 8). The data are also
very noisy. The very deviant observations and the noise are not due to
instrumental problems since observations for other stars are stable (see
Section~\ref{section-var-tests}) and some of the observations for
MWC~1080 consist of $2\;$--$\;5$ individual measurements which are all
consistent with one another.

We have also obtained data at the OPdM (France) in October 1996, using a
$V$ filter and a $10\arcsec$ aperture hole. These observations are
presented in Table~\ref{Tab-Pic-V}, along with data from the literature
and OMM. The $V$ observations from OMM and OPdM are presented in
Figure~\ref{Fig-MWC_V}. Observations were also obtained at OMM on 3
occasions at other wavelengths to produce curves of the linear
polarization as a function of wavelength (see Table~\ref{Tab-P_lambda}).

\subsubsection{Interstellar and intrinsic polarization\label{section-is-pol}}
% Notes to myself: 
% Mathewson et al 1978 catalog: available online only at CDS, catalog 
% II/34A/catalog (the 1978 source is a note in the Info. Bull. of CDS, the
% original data from 1970 is a Memoirs of the RAS), and I don't know the
% wavelength (probably unfiltered). The output for the catalog search is
% a text file in the directory MATTHEWSON; there is also an analysis
% program (P. Bastien) with selected stars, and graph, in the directory
% ISPNS_PROG/ (see procedure above). 

An inspection of the Mathewson et al.  (1978) catalog shows that
MWC~1080 is in a region where the position angles are well aligned at
$73\;$--$\;74\arcdeg$ (simple and weighted averages of the position
angles for 51 stars found within 4\arcdeg\ of MWC~1080 and having
similar distance modulus), with a wide distribution of polarization
values that go up to 5\% (see Figure~\ref{Fig-pol_IS}). This indicates
that the interstellar polarization is strong in this region, and we
estimate it to be, based on polarization for neighboring stars, $2.21
\pm 0.13$\% at 73\arcdeg\  in the $V$ band, similar to the estimation by
Hillenbrand et al. (1992) of 2.58\% at 71\arcdeg\ in the $V$ band, but
perpendicular to the Garrison \& Anderson (1978) value of 2.00\% at
168\arcdeg. We believe, based on Hillenbrand et al.'s (1992) and our
analysis of the Mathewson et al. (1978) catalog, that the interstellar
polarization has a position angle of $\approx 75\arcdeg$ and not
$\approx 165\arcdeg$.

Using our estimate of the IS polarization, 2.21\% at 73\arcdeg\  in the
$V$ band, and Serkowski's law for interstellar polarization (Serkowski,
Mathewson, \& Ford 1975), we find an interstellar polarization value of
1.95 \% at $73\arcdeg$ at $7660\;$\AA. We have assumed that
$\lambda_{\rm max}=5500\;$\AA\ for Serkowski's law (a typical value, and
also a mean for 3 stars found near MWC~1080 in the Mathewson et al.
(1978) catalog, and a value compatible with our wavelength-dependent
data) and that the position angle of the interstellar polarization is
constant as a function of wavelength.

From our data, MWC~1080's average polarization is 1.60\% at 81.6\arcdeg\
at $7660\;$\AA\ ($N=62$), and 1.96\% at 76.4\arcdeg\ at $V$ ($N=7$, data
from OPdM and OMM only). Nine stars are found within 1\arcdeg\ of
MWC~1080. Two, at about half a degree, show polarization levels of
1.36\% and 1.54\%, with position angles of 62 and 63\arcdeg, which
indicates that a significant part of MWC~1080's polarization is of
interstellar origin.

A proper subtraction of the interstellar polarization from the observed
polarization (using the Stokes parameters $Q=P\;\cos 2\theta$ and
$U=P\;\sin 2\theta$ and not a usual vectorial subtraction as was
apparently done in the Hillenbrand et al. (1992) paper, in particular in
their table~2) gives estimates of the average intrinsic polarization for
MWC~1080: 0.6\% at 139\arcdeg\ at $7660\;$\AA, and 0.35\% at 142\arcdeg\
in the $V$ band. A proper subtraction taking the Hillenbrand et al.
(1992) $V$ data would give $0.84\%$ at 152\arcdeg; Garrison \& Anderson
(1978) report a much higher and perpendicular polarization, 3.84\% at
76\arcdeg\ but we believe this is due to an inappropriate evaluation of
the interstellar polarization.

The position angles $\approx\,140\arcdeg$ trace the axis of symmetry of
the system and are at about 90\arcdeg\ with respect to the position
angle ($\approx 60\arcdeg$) of the outflow cavity (Poetzel, Mundt, \&
Ray 1992), which is what is usually observed for PMS stars (Bastien
1988). We then confirm the Poetzel et al. (1992) scenario where the
elongated structure they saw in the continuum at $7018\;$\AA\ is made of
the innermost and brightest part of a reflection nebula, elongated in
the outflow direction, and perpendicular to the polarization vector.

The significant intrinsic polarization indicates that the circumstellar
or circumbinary material around this binary is arranged in an asymmetric
geometry, for example, in a disk rather than a spherical shell.

Two other facts point to an intrinsic polarization component in the
observed polarization for MWC~1080. First, MWC~1080 shows clear periodic
variations (see below) that cannot be of interstellar origin. Second,
the position angle of the polarization varies as a function of
wavelength (see below). Observations of polarized standard stars (whose
polarization is of interstellar origin) by Schmidt, Elston, \& Lupie
(1992) show that the maximum observed rotation of the position angle for
the wavelength range covered by our observations (4350--$8580\;$\AA) is
$\approx 0.25 \arcdeg$, much less than the $\sim 10 \arcdeg$ rotation
seen for MWC~1080.

As discussed by Dolan \& Tapia (Dolan \& Tapia 1986 and references
cited), a wavelength-dependent position angle could in theory be
attributed to multiple interstellar clouds containing different grain
sizes magnetically aligned in different directions, or to an
interstellar cloud where there is a continuous rotation of the grains'
orientation, but, once again, the variability argument points to the
presence of intrinsic polarization.

%%%%%%%%%%%%%%%%%%%%%%%%%
\subsubsection{Variability analysis}
Already from Figure~\ref{Fig-MWC1080_all}, it is obvious that MWC~1080's
polarization is variable. The detailed results of the variability tests
for the $7660\;$\AA\ wavelength and the $V$ filter will be presented in
a future paper, along with analysis for our other binaries. The analysis
shows that MWC~1080 was clearly variable polarimetrically between August
1995 and June 1999, in both filters, with indications of variability on
time scales of days and months.

The average polarization at $7660\;$\AA\ for the period 1995--1999 is
1.6\% at 82\arcdeg, whereas unpublished data taken at the OMM on Jan. 15
1988 in the same filter and aperture hole show a significantly higher
polarization value, $1.98\;\pm\;0.04$\%, but with a similar position
angle, 81.7\arcdeg. This points to long-term (years) variability.

Data for the $V$ filter show that there might be a significant
difference between observations taken before and after October 1973, as
data from 1973 (Vrba et al. 1975) show a higher polarization and
(slightly) different position angle. After 1973, the polarization and
its angle seem to have stabilized, although there is still evidence of
variability. In particular, the four observations taken at OPdM on 4
consecutive nights clearly show variability. Once again, this indicates
that variability is present with different time scales.

On some nights (6), it was possible to get 3 observations within about 5
hours. On many occasions, a set of 3 points shows a systematic increase
or decrease in polarization and/or position angle, usually by a
significant amount, showing that the variations are not random. On other
occasions, the position angle stayed constant. This indicates that
variability is present on time scales of hours, as well as days,
months, and years.

The variations in position angle are confined to within $\pm\;5$\arcdeg\ 
of the mean, which indicates that polarization variations are probably
not due to a drastic change in the geometry of the circumstellar
material, but rather to modifications in the density.

\subsubsection{Periodic polarimetric variations}
Figure~\ref{Fig-MWC1080_all} shows all the $7660\;$\AA\ observations,
and Figure~\ref{Fig-MWC1080_bin20} shows binned data, where the orbital
phase has been divided in equal bins and the polarization data weight
averaged; the error bars in Figure~\ref{Fig-MWC1080_bin20} are simple
averages of the error bars of the data in each bin. Clear periodic
variations with amplitudes in $Q$ and $U$ $\approx 0.1-0.2$\% are seen,
especially in polarization angle, but are not fitted well by the simple
$1\lambda$ and $2\lambda$ sinusoidal curves; the coefficients for this
fit are (Equations~\ref{p3-eq-qfit} and \ref{p3-eq-ufit}):

\parbox{3in}{
\begin{eqnarray}
q_0 &=& -1.5229 \pm 0.0011 \\
q_1 &=&  0.0278 \pm 0.0014\\
q_2 &=& -0.0260 \pm 0.0016\\
q_3 &=& -0.0404 \pm 0.0015\\
q_4 &=&  0.0430 \pm 0.0015
\end{eqnarray} }
\hfill
\parbox{3in}{
\begin{eqnarray}
u_0 &=&  0.4536 \pm 0.0037\\
u_1 &=& -0.0162 \pm 0.0049\\
u_2 &=& -0.0141 \pm 0.0053\\
u_3 &=& -0.0012 \pm 0.0051\\
u_4 &=& -0.0587 \pm 0.0051 
\end{eqnarray} }

Adding $3\lambda$ and $4\lambda$ harmonics does a
better job (see Figure~\ref{Fig-MWC1080_fit4}), but the fit is still not
adequate; the coefficients for that fit are:

\parbox{3in}{
\begin{eqnarray}
q_0 &=& -1.5189 \pm 0.0004 \\
q_1 &=&  0.0286 \pm 0.0005\\
q_2 &=& -0.0309 \pm 0.0005\\
q_3 &=& -0.0438 \pm 0.0005\\
q_4 &=&  0.0376 \pm 0.0005\\
q_5 &=&  0.0345 \pm 0.0005\\
q_6 &=&  0.0083 \pm 0.0005\\
q_7 &=& -0.0145 \pm 0.0005\\
q_8 &=&  0.0118 \pm 0.0005
\end{eqnarray} }
\hfill
\parbox{3in}{
\begin{eqnarray}
u_0 &=&  0.4557 \pm 0.0037\\
u_1 &=& -0.0134 \pm 0.0051\\
u_2 &=& -0.0162 \pm 0.0056\\
u_3 &=& -0.0045 \pm 0.0053\\
u_4 &=& -0.0625 \pm 0.0053\\
u_5 &=&  0.0148 \pm 0.0053\\
u_6 &=&  0.0169 \pm 0.0052\\
u_7 &=& -0.0021 \pm 0.0050\\
u_8 &=& -0.0038 \pm 0.0055
\end{eqnarray} }

The level of the the polarimetric variations could point to a
circumstellar disk rather than a circumbinary disk, as we have shown in
Paper~II that the former configuration is more favorable to high
amplitude variations. However, if we assume both stars in MWC~1080 have
masses of $\sim 10\;$M$_{\sun}$, the orbital period yields a separation
of $0.1\;$AU between each star; if the radii are $\sim 5\;$R$_{\sun}$,
or about $0.025\;$AU, the photospheres are separated by only $0.05\;$AU,
which does not leave much space for dusty circumstellar material (which
might then just evaporate due to the proximity of the photospheres).

In the observed polarimetric curves, there are minima of polarization
and maxima in position angle near phases 0.0 and 0.5. When the
interstellar polarization, which is almost perpendicular to the observed
one, is removed from the observed polarization, the intrinsic
polarization now shows a maxima near phases 0.0 and 0.5, close to the
position of the primary and secondary eclipses (0.0 and 0.55 (Shevchenko
et al.  1994)). Both $7660\;$\AA\ and $V$ observations show a sudden
increase in intrinsic polarization around phase 0.5 (see
Figures~\ref{Fig-MWC_V} and \ref{Fig-MWC1080_bin20}). Since this change
is seen in both filters, it must be related to the eclipses. An increase
in polarization during an eclipse is to be expected since shielding the
unpolarized light from the star will increase the amount of the
scattered (and polarized) light.  Increase of polarization at phase 0.0
had already been detected in the $U$, $B$, and $V$ bands by Shevchenko
et al. (1994).

A prediction of our Paper I was that as the orbital eccentricity
increases, the periodic variations start from pure double-periodic to
include stronger and stronger single-periodic variations.  The
eccentricity for MWC~1080 is not known but is estimated to be between
0.2 and 0.5 (Grankin et al.  1992); the strength of double-periodic
variations, which have amplitudes between 1.4 and 2.5 times those of the
single-periodic one, would favor an eccentricity nearer to 0.2 than to
0.5, but the polarimetric curves can not be formally inverted to find the
orbital eccentricity.

In addition, non-periodic variations introduce noise that contributes,
sometimes substantially, to both harmonics, so that such a
straightforward association between eccentricity and presence of
$1\lambda$ variations cannot be made. We have shown in Papers~I and
II that factors other than eccentricity (time-varying optical depth, non
equal luminosity stars, asymmetric geometries, scattering on dust
grains) can also introduce such single-periodic variations.

The PDM finds periods of 1.1 and $2.5\;$d when we look at a subset of
the data taken over a few months only, and the LNP shows a peak at
$1.6\;$d which becomes significant (more than 97\% chance that the data
do not come from random Gaussian noise) when we look at the binned
data. The periods of $1.6\;$d and $2.5\;$d could be related to the
orbital period of $2.88\;$d.

%%%%%%%%%%%%%%%%%%%%%%%%%
\subsubsection{Polarization as a function of wavelength}
On three occasions, in 1996, 1997, and 1998, data were obtained in order
to get $P(\lambda)$, the polarization as a function of wavelength. Data
are presented in Table~\ref{Tab-P_lambda} and
Figure~\ref{Fig-P_lambda}. The polarization and its position angle are
clearly variable as a function of wavelength and time. The observed
position angle generally decreases towards shorter wavelengths, by up to
10\arcdeg\ between $4350\;$\AA\ and $8580\;$\AA. Changes of up to
10\arcdeg\ are also seen from year to year. Variations like those have
already been observed for T~Tauri stars (Bastien 1981).

The change of polarization angle as a function of wavelength is mostly
an intrinsic variation, since the interstellar polarization shows minute
variations as a function of wavelength (once again, see Schmidt, Elston,
\& Lupie 1992). Also, a study of the variability of polarized standard
stars (Bastien et al. 1988), where the variability might due to IS
polarization variability, shows variations $\lesssim 5\arcdeg$, also
below those seen for MWC~1080. We therefore believe that IS polarization
variability would not be able to explain the change of position
angle as a function of wavelength for MWC~1080.

As data taken during the same night at the same wavelength show,
significant variations might have occurred during the time the
$P(\lambda)$ was measured. Nonetheless, the 3 curves seem to be
relatively consistent from one data set to the other. The polarization
has a general bell shape usually associated with interstellar
polarization and the Serkowski law (Serkowski, Mathewson, \& Ford
1975). In particular, this interstellar curve has a peak at $\approx
5500\;$\AA, near the peak of MWC~1080's curve, which indicates, as
already stated, that MWC~1080's polarization is in great part of
interstellar origin.

Using our estimate of the IS polarization for the $V$ band and
Serkowski's law, we can compute the IS polarization for the observed
wavelengths, and subtract it to find the intrinsic $P(\lambda)$ (see
Figure~\ref{Fig-MWC_instrinsic}). Removal of the IS polarization moves
the peak of polarization to a redder wavelength around $7000\;$\AA.

Using the numerical code presented in Papers~I and II, we have computed
a few $P(\lambda)$ curves for different grain composition and
sizes. Even though the number of free parameters is too high to uniquely
determine the composition and size of the grains from the observed
$P(\lambda)$ curve, these curves can nonetheless give some insight into
the environment for MWC~1080.  Figure~\ref{Fig-P_lambda_AS} presents 3
$P(\lambda)$ curves for astronomical silicate grains of radii
$0.02\;\micron$, $0.1\;\micron$, and $0.2\;\micron$. It is immediately
seen that the size of the grains has a profound impact on the morphology
of that curve.

From our simulations, the circumbinary disk configuration does not
produce variations of the polarization angle as a function of
wavelength, but the circumstellar disk does, for certain orbital
phases. The variations are small, 3--4\arcdeg, but the dependence can be
either increasing or decreasing toward the blue, depending on the
orbital phase. Since MWC~1080 does show a variable polarization angle,
it points to a circumstellar disk, or at least, to the presence of matter
close to the stars.

Qualitatively, then, the polarization for MWC~1080 could be reproduced
by astronomical silicate grains of small sizes, between $0.02\;\micron$
and $0.1\;\micron$, located close to the stars, in an asymmetric
configuration. 

\section{Photometry}
Photometric observations of the irregular variable MWC~1080 show
periodic variations with normal light variations of 0.16 magnitudes in
$V$, along with irregular variability with $\Delta V = 0.3$ magnitudes,
which introduces scatter in the periodic variations (Grankin et
al. 1992). An inspection of a catalog of $UBVRI$ photometry of T~Tauri
stars (Herbst et al. 1994) with its on-line database\footnote{Available
at {\tt{ftp://www.astro.wesleyan.edu/pub/ttauri/}}} containing $\approx
1600$ observations for this binary also reveals that the average
magnitude for MWC~1080 can change from one epoch to
another. Peak-to-peak variations are $0.2\;$--$\;0.3$ magnitudes in $V$,
whereas the scatter is of the same order. This stochastic photometric
variability is probably related to the stochastic polarimetric
variability observed here.

We have tried to cross-correlate photometric observations taken from
that database with our polarimetric observations, and find data
obtained quasi-simultaneously. Due to the physical locations of the
observatories, the closest observations are within $0.25\;$--$\;0.5$ days
of each other, which is a significant portion of the orbital
period. Nonetheless, we found about half a dozen quasi-simultaneous
observations, which do not show any clear correlation between brightness
of the star and its intrinsic polarization.

\section{Orbital inclination}
One of the goals of these observations is to find the orbital
inclination for this binary system. Since it is an eclipsing binary, the
orbital inclination is known to be close to 90\arcdeg, but the light
curve also suggests some small departure from an edge-on configuration
(Grankin et al. 1992). Since the orbit might be significantly eccentric
($0.2<e<0.5$, Grankin et al. 1992), we should consider the results of
the BME formalism using both orders (Equations~\ref{EQ-iO1-p3} and
\ref{EQ-iO2-p3}), which are $i(O1)=56\;\pm\;8\arcdeg$ and
$i(O2)=102\;\pm\;2\arcdeg$.

If the orbital eccentricity is $e \approx 0.2\;$--$\;0.3$, both orders
should give about the same inclination, which is not the case. If the 
orbital eccentricity is higher than $\approx 0.3$, only the first-order
results should be used ($i(O1)=56\;\pm\;8\arcdeg$), but this result is
not compatible with an eclipsing system, although the separation
between the two components is small.

In Paper~I, we showed that the amplitude of the polarimetric variations
and the stochastic noise (or, scatter of the observations about the mean
polarimetric variations) must be considered first in order to see if the
orbital inclination found from Equations~\ref{EQ-iO1-p3} and
\ref{EQ-iO2-p3} is meaningful or not. According to our noise criteria
(found after adding different levels of noise in the polarimetric curves
produced by our numerical simulations; see Paper~I for more details),
the noise present in MWC~1080's data is too high (the scatter is
$\approx 0.08$\% in polarization, or 50\% of the amplitude of the
variations) to allow the BME formalism to be used. Even applying the
same analysis to the binned data, less numerous but showing much less
scatter, gives the same conclusion.

The data have to be of exquisite quality (high number of observations,
clear variations, not much stochastic noise) in order for the
inclination found with the BME formalism to be meaningful.  In the case of
hot stars, the variations are usually of higher amplitude (by a factor
of $\sim 3$) with less scatter from intrinsic non-periodic variations,
so that in those cases, the noise criteria indicates the BME results are
meaningful.

% Drissen et al 1989, ApJ, 343, 426 - EZ CMa - delta_P=0.6%, delta_P=0.3%
%         some scatter
% Cellone et al 1996, Rev MexAA, 5, 123 - WR binary - delta_Q=0.3% with
% scatter, delta_U=0.6% with scatter

%%%%%%%%%%%%%%%%%%%%%%%%%%%%%%%%%%%%%%%%%%%%%%%%%%%%%%%%%%%%%%%%%%%%%%%%
% DISCUSSION
\section{Summary and conclusions}
We have presented polarimetric observations of a young massive
short-period binary star of the Herbig~Ae/Be type, MWC~1080. The mean
polarization at $7660\;$\AA\ is 1.60\% at 81.6\arcdeg\  (or 0.63\% at
139\arcdeg\ if an estimate of the interstellar polarization is
subtracted) and 1.96\% at 76.4\arcdeg\ in the $V$ band. The significant
intrinsic polarization points to an asymmetric geometry of the
circumstellar or circumbinary environment. The 139\arcdeg\ intrinsic
position angle traces the axis of symmetry of the system and is
perpendicular to the position angle of the outflow cavity ($\approx
60\arcdeg$).

Although the interstellar polarization is very strong in this region,
MWC~1080's polarization cannot be of interstellar origin only, for 3
reasons. First, the estimated interstellar polarization for this region
is different than the observed one for MWC~1080. Second, the significant
variability of the polarization as a function of time (especially on
time scales of hours) cannot be explained by modifications in the
interstellar medium. Third, the dependence of the position angle on the
wavelength along with its variability cannot be attributed to the
interstellar medium.

This polarization is clearly variable, at all wavelengths, and on time
scales of hours, days, months, and years; this is not a surprise and was
expected since many PMS stars, of the Herbig~Ae/Be or T~Tauri type,
show such variations. In addition to that stochastic variability, there
are periodic variations due to the orbital motion of the stars in their
dusty environment. Although stochastic (non-periodic) polarimetric
variations introduce noise and scatter around the mean variations,
binning the numerous observations reveals clear periodic polarimetric
variations, which follow the known orbital period. The periodic
polarimetric variations of this binary star are the first phased-locked
ones detected for a PMS binary star that the authors are aware of.

The amplitude of the variations point to a circumstellar disk rather
than a circumbinary one, since, according to our numerical simulations,
the second configuration does not produce significant
variations. However, the stars are probably so close to one another that
there is not much room left for circumstellar material.

The variations are not simply double-periodic (as produced by a simple
model of 2 equal mass stars in a circular orbit, at the center of an
axisymmetric circumbinary envelope made of electrons), but include
single-periodic and higher order variations. The presence of
single-periodic variations could be due to non equal mass stars
(although photometric observations point to equal mass or similar mass
stars), the presence of dust grains, an asymmetric configuration of the
circumstellar or circumbinary material, or the eccentricity of the orbit
(which is thought to be between 0.2 and 0.5). The available data are for
the moment insufficient to say what is the predominant cause of those
single-periodic variations, or to find the geometry or orbital
eccentricity of this system.

MWC~1080 is an eclipsing binary with primary and secondary eclipses
occurring at phases 0.0 and 0.55. The signatures of the eclipses are seen
in the polarimetric observations. The primary eclipse coincides with 
maxima in intrinsic polarization and position angle in the
$7660\;$\AA\ data. In addition, both $7660\;$\AA\ and $V$
observations show a sudden increase in polarization around phase 0.5
where the secondary eclipse is located. 

Most of the observations have been obtained at a wavelength of
$7660\;$\AA, but we have obtained additional $V$ band observations, and
also observations as a function of wavelength on 3 occasions. The
$P(\lambda)$ curve has a bell-shape morphology which changes on a time
scale of a year, with a maximum polarization at about $5500\;$\AA,
typical of the interstellar polarization; when the estimated
interstellar polarization is removed, the peak in the $P(\lambda)$ curve
is displaced towards $\approx 7000\;$\AA. The position angle also
depends on the wavelength, with a rotation of $\sim 10\arcdeg$ over
$\sim 4000\;$\AA, and on time. Both MWC~1080's polarization and position
angle are thus dependent on time and wavelength.

The intrinsic polarization and intrinsic $P(\lambda)$ curves are
consistent with the presence of dusty material in the circumstellar
environment. Although it is not possible to uniquely determine the
composition and size of the dust grains, our observations would be
compatible with astronomical silicate grains with sizes in the range
$0.02\;$--$\;0.1$\micron.

Photometry for this irregular variable star shows stochastic variations
that introduce significant scatter in the photometric light curve, a
characteristic also seen in our polarimetric data. No clear correlation
between brightness and polarization was found.

The stochastic polarimetric variations mentioned before introduce so
much scatter that the orbital inclination obtained with the BME
formalism is unfortunately not meaningful. This example shows that
polarimetric observations have to be of exquisite quality (high number
of observations, clear variations, not much stochastic noise) in order
to find the orbital inclination.

Data on about 2 dozen other spectroscopic binary young stars will be
presented in future papers.

%%%%%%%%%%%%%%%%%%%%%%%%%%%%%%%%%%%%%%%%%%%%%%%%%%%%%%%%%%%%%%%%%%%%%%%%
% ACKNOWLEDGMENTS
\acknowledgments
N. M. would like to thank the Conseil de Recherche en Sciences
Naturelles et G\'enie of Canada, the Fonds pour la Formation de
Chercheurs et l'Aide \`a la Recherche of the province of Qu\'ebec, the  
Facult\'e des Etudes Sup\'erieures and the D\'epartement de physique
of Universit\'e de Montr\'eal for scholarships, and P. B. for
financial support. We would like to thank the Conseil de Recherche en 
Sciences Naturelles et G\'enie of Canada for supporting this research.

The authors gratefully acknowledge financial support from Universit\'e
Joseph Fourier, Grenoble (1994 BQR grant B 644 R1), the Laboratoire
d'Astrophysique de l'Observatoire de Grenoble, and the Observatoire
Midi-Pyr\'en\'ees for upgrade and maintenance of the STERENN polarimeter.

We would like to thank F. M\'enard for taking the Pic-du-Midi
observations, and E. Magnier for reading the manuscript.

%%%%%%%%%%%%%%%%%%%%%%%%%%%%%%%%%%%%%%%%%%%%%%%%%%%%%%%%%%%%%%%%%%%%%%%%
% APPENDICES

%%%%%%%%%%%%%%%%%%%%%%%%%%%%%%%%%%%%%%%%%%%%%%%%%%%%%%%%%%%%%%%%%%%%%%%%
% BIBLIOGRAPHY (new page)

%%%%%%%%%%%%%%%%%%%%%%%%%%%%%%%%%%%%%%%%%%%%%%%%%%%%%%%%%%%%%%%%%%%%%%%%
% FIGURE CAPTIONS (new page)
\newpage

\scalebox{0.75}{\includegraphics{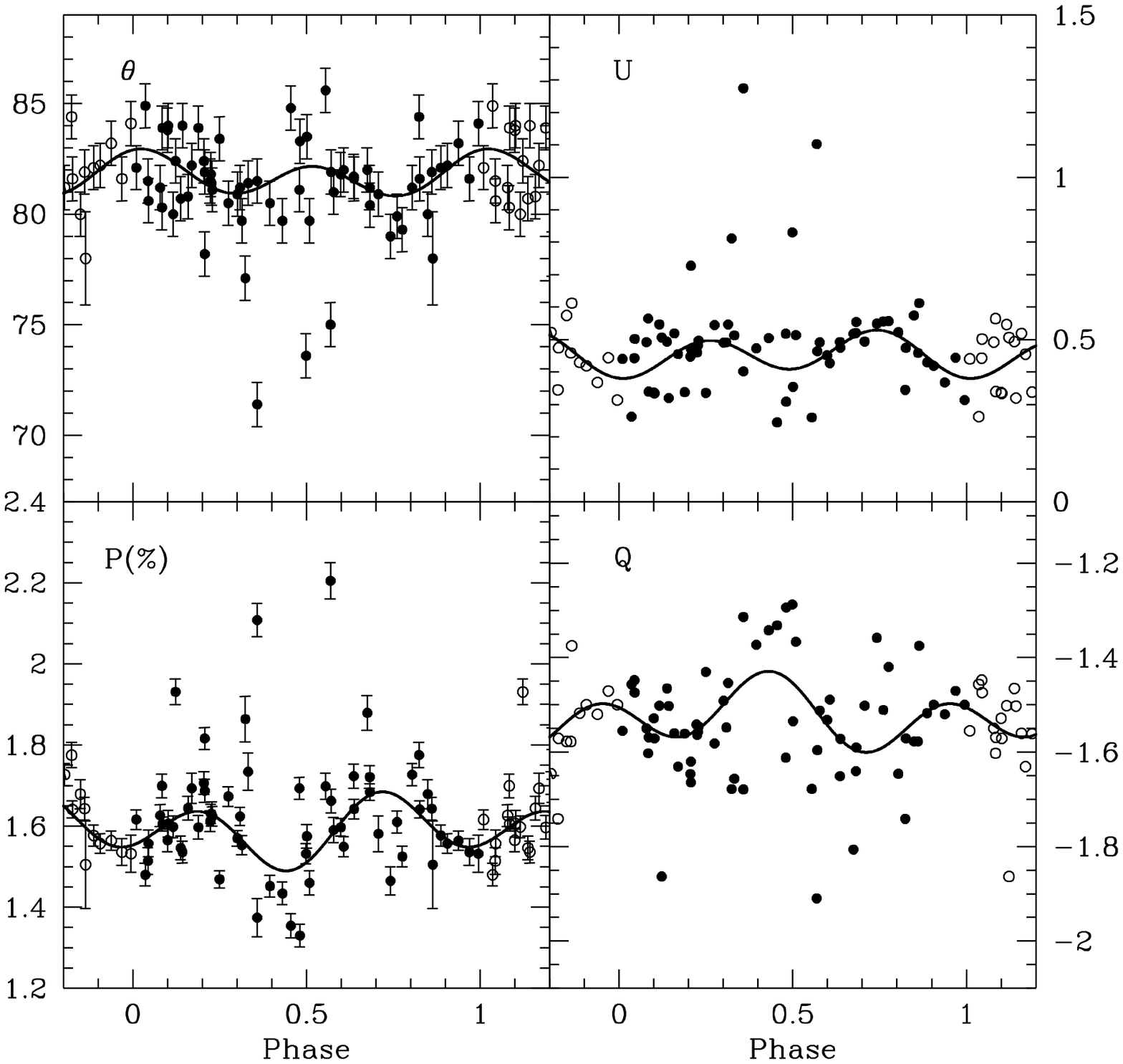}}
\figcaption[Manset3_v2.fig1.ps]{Complete set of polarimetric observations of
MWC~1080, taken at $7660\;$\AA. The ephemeris used is from Grankin et
al.  (1992): 2445607.374 $+$ 2.886926$E$. The open circles are
duplicates of filled circles and were added for clarity. A few
observations are clearly away from the average curve, in both
polarization and position angle, and are due to intrinsic
variations. The solid line is the Fourier fit made according
Equations~\ref{p3-eq-qfit} and \ref{p3-eq-ufit}. MWC~1080 is an
eclipsing binary; the primary eclipse occurs at phase 0.0, and the
secondary at phase 0.55.\label{Fig-MWC1080_all}}

\newpage
\scalebox{0.75}{\includegraphics{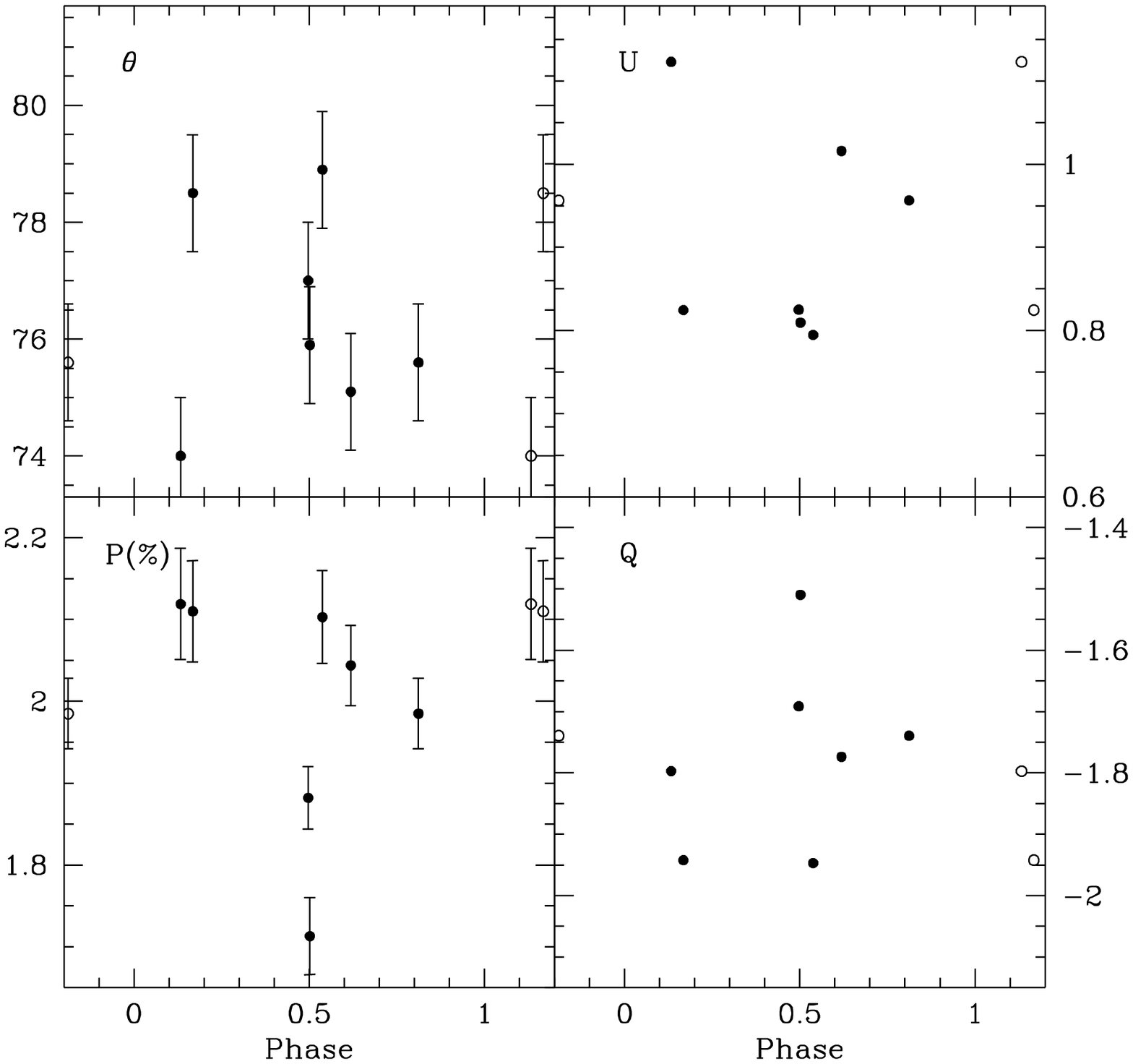}}
\figcaption[Manset3_v2.fig2.ps]{Polarimetric observations of MWC~1080, taken in
the $V$ filter at Pic-du-Midi and Mont M\'egantic observatories. The
ephemeris used is from Grankin et al.  (1992). \label{Fig-MWC_V}}

\newpage
\scalebox{0.75}{\includegraphics{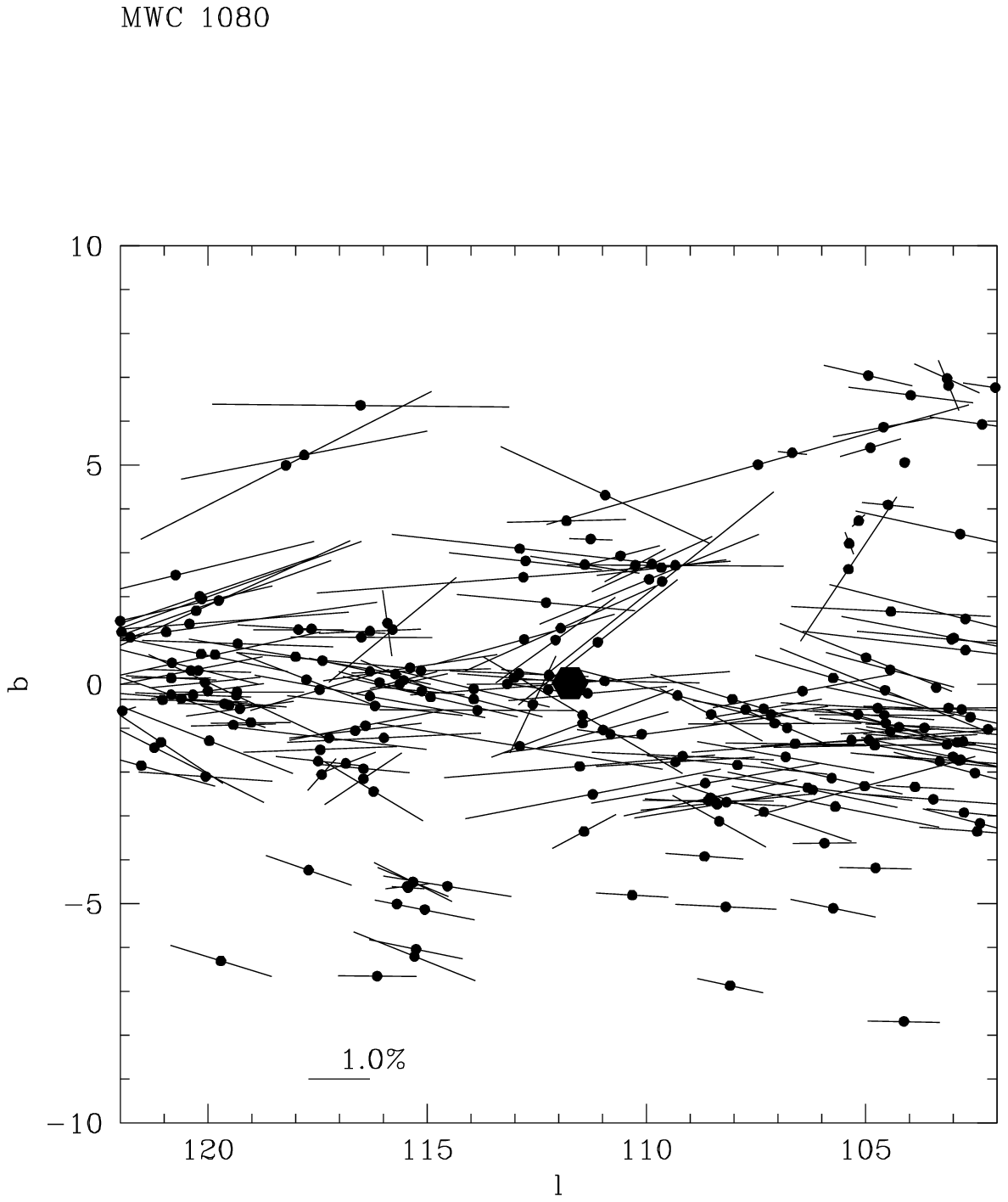}}
\figcaption[Manset3_astroph.fig3.ps]{Interstellar polarization in the vicinity
of MWC~1080, shown at the center of the field. Data are from the
Mathewson et al.  (1978) catalog. \label{Fig-pol_IS}}

\newpage
\scalebox{0.75}{\includegraphics{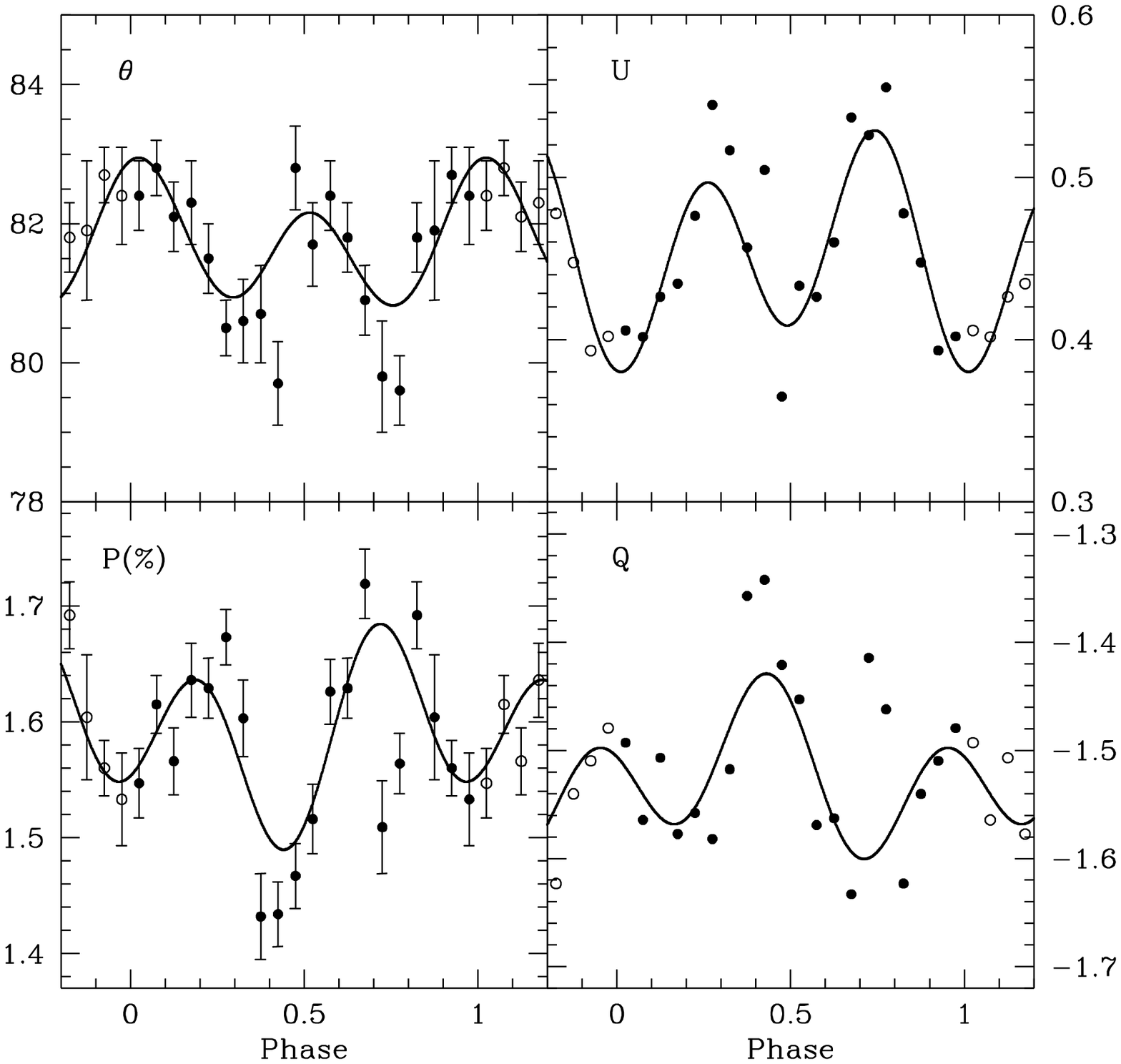}}
\figcaption[Manset3_v2.fig4.ps]{Binned polarimetric observations of
MWC~1080. The most deviant points were removed before binning the data
into 20 bins. Periodic variations are clearly seen. Note the steep
change in polarization near phase 0.5. \label{Fig-MWC1080_bin20}}

\newpage
\scalebox{0.75}{\includegraphics{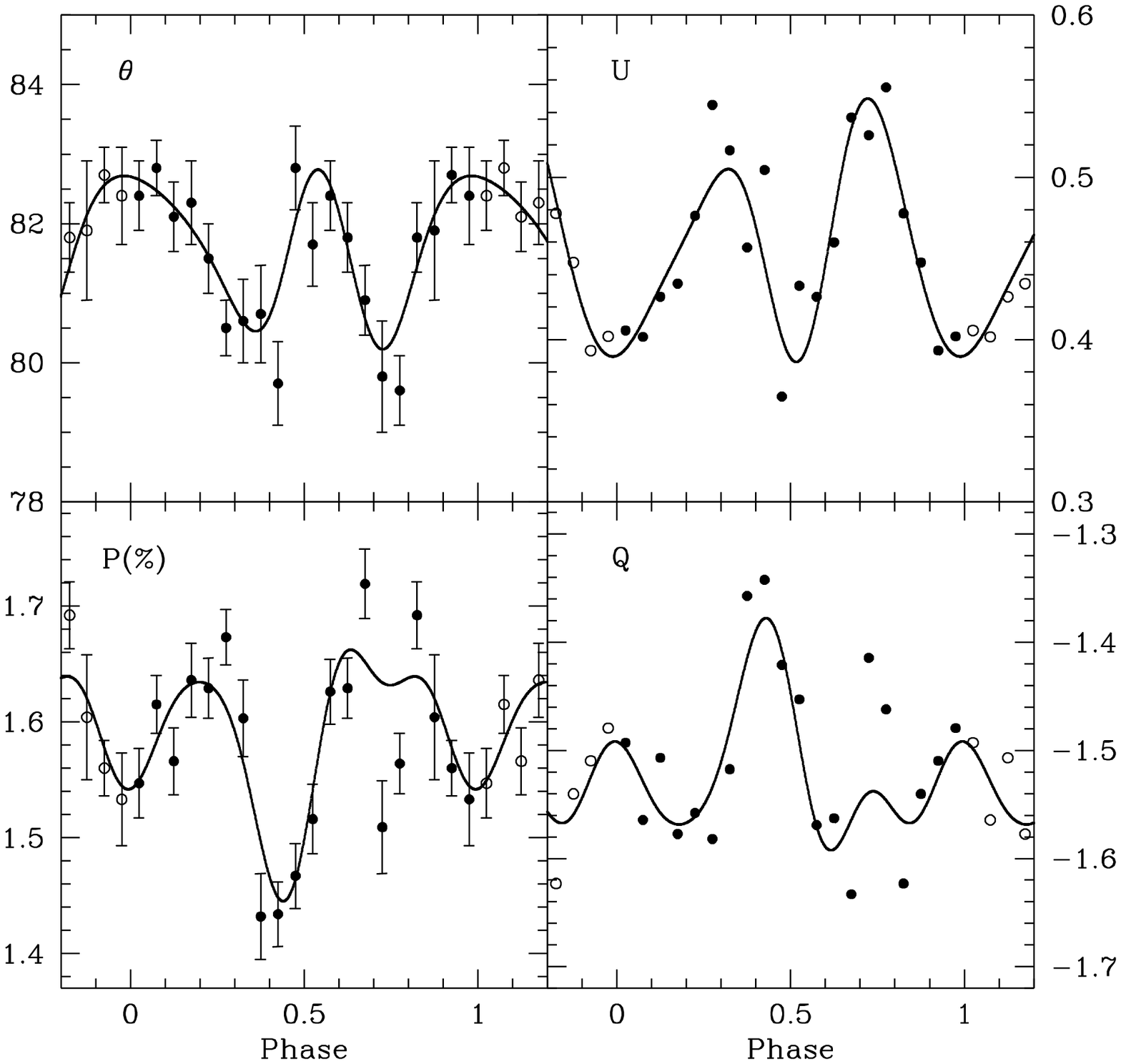}}
\figcaption[Manset3_v2.fig5.ps]{Binned polarimetric observations of
MWC~1080, with an order 4 fit, showing that harmonics higher than second
are present in the data. \label{Fig-MWC1080_fit4}}

\newpage
\scalebox{0.75}{\includegraphics{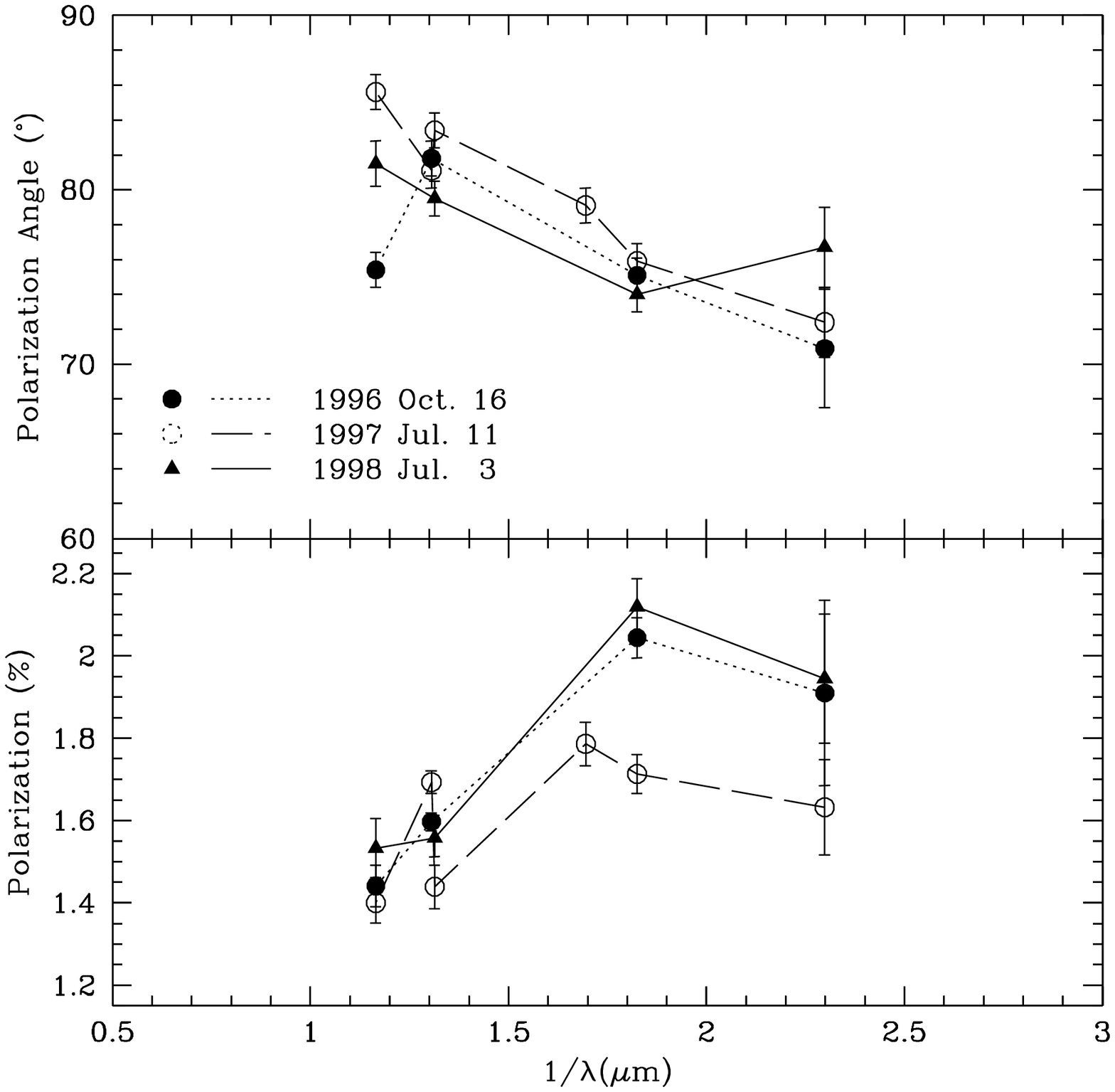}}
\figcaption[Manset3_v2.fig6.ps]{Observed polarization (intrinsic +
interstellar) for MWC~1080 as a function of wavelength, for 3 different
dates. \label{Fig-P_lambda}}

\newpage
\scalebox{0.75}{\includegraphics{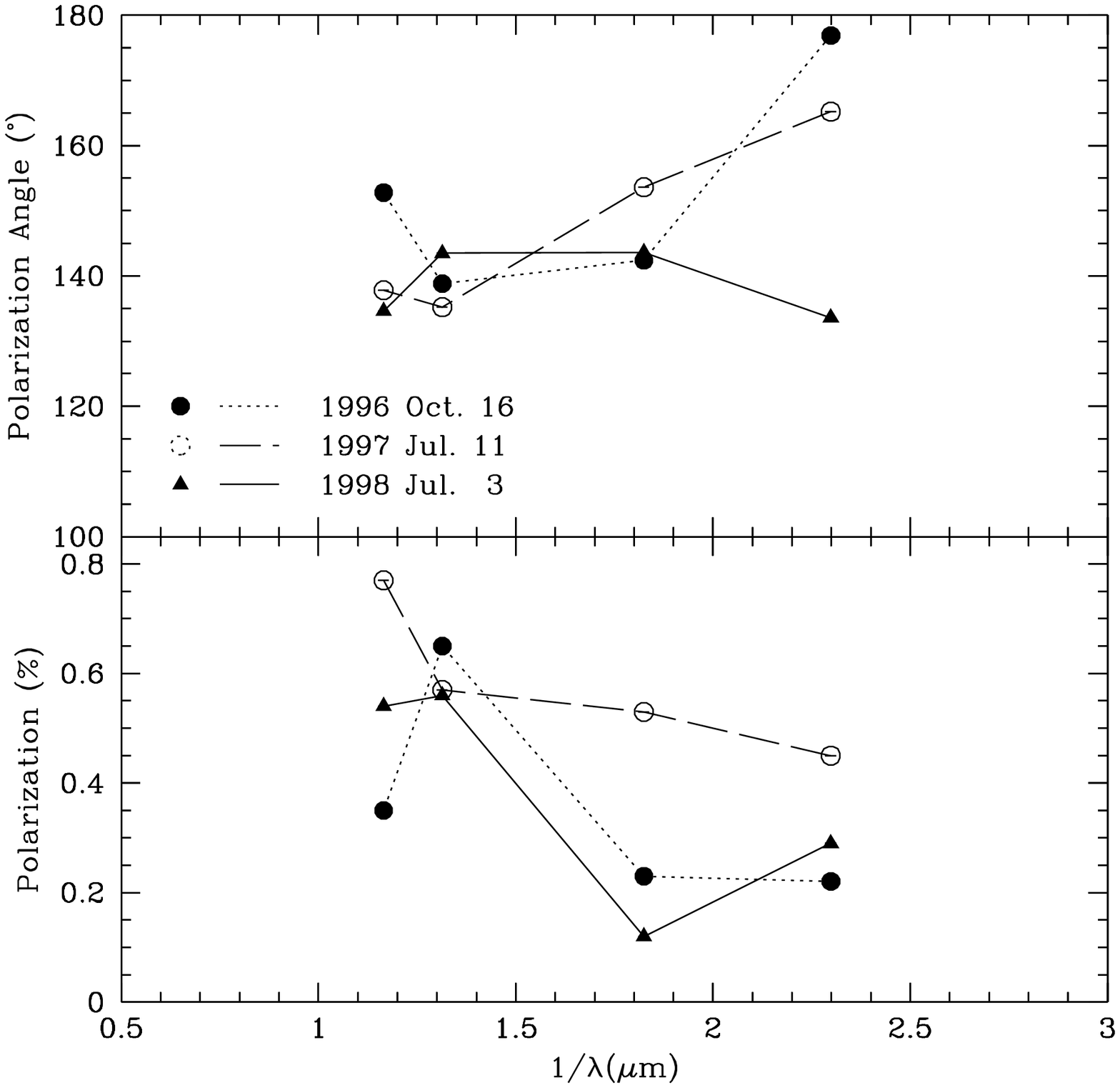}}
\figcaption[Manset3_v2.fig7.ps]{Intrinsic polarization for MWC~1080,
after removal of the IS polarization. See text for more
details. \label{Fig-MWC_instrinsic}}

\newpage
\scalebox{0.75}{\includegraphics{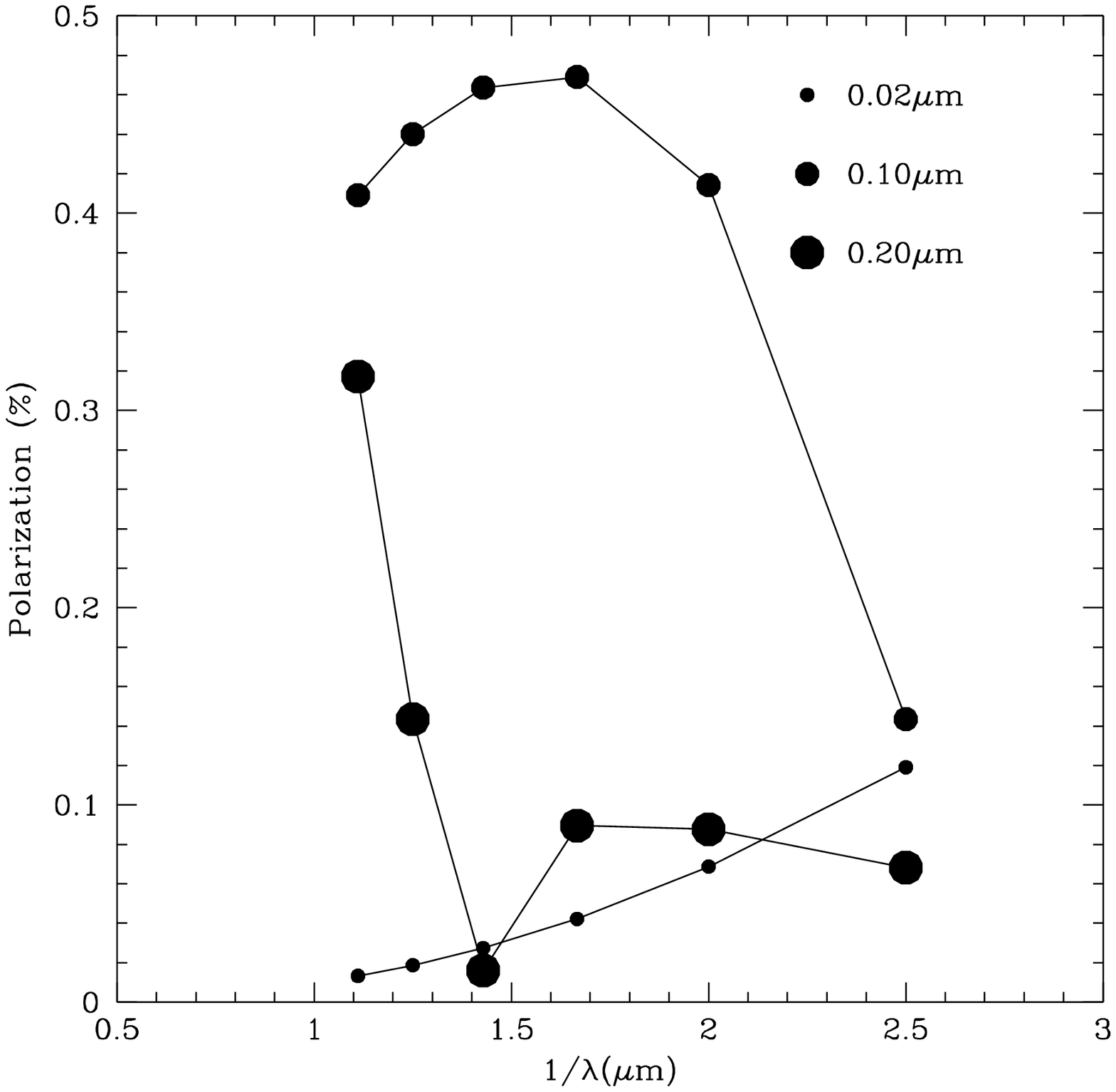}}
\figcaption[Manset3_v2.fig8.ps]{Polarization as a function of wavelength for
astronomical silicate grains of radii $0.02\;\micron$, $0.10\;\micron$, 
and $0.20\;\micron$. Compare with the previous
figure. \label{Fig-P_lambda_AS}}

%%%%%%%%%%%%%%%%%%%%%%%%%%%%%%%%%%%%%%%%%%%%%%%%%%%%%%%%%%%%%%%%%%%%%%%%
% TABLES (new page)

\newpage
% ---------------------- ALL DATA ----(66)------------------------------
\begin{deluxetable}{lcccccc}
\tablewidth{0pt}
\tablecaption{Polarization data for MWC~1080, $7660\;$\AA, Observatoire du Mont
M\'egantic \label{Tab-data}}
\tablehead{
\colhead{UT Date} & \colhead{JD} & \colhead{Phase\tablenotemark{1}} & 
\colhead{P(\%)} & \colhead{$\sigma(P)$} &
\colhead{$\theta(\arcdeg)$} & \colhead{$\sigma(\theta)$}\\
\colhead{} & 2400000.0$+$ & \colhead{} &
\colhead{} & \colhead{} &
\colhead{} & \colhead{}}
\startdata
1995 Aug 17 & 49946.824 & 0.138 & 1.546 & 0.029 & 80.7 &  1.0\\  
1995 Aug 19 & 49948.809 & 0.826 & 1.641 & 0.021 & 81.6 &  1.0\\  
1995 Aug 26 & 49955.683 & 0.207 & 1.687 & 0.029 & 81.9 &  1.0\\  
1995 Aug 28 & 49957.646 & 0.887 & 1.577 & 0.026 & 82.1 &  1.0\\  
1995 Aug 29 & 49958.617 & 0.223 & 1.612 & 0.026 & 81.5 &  1.0\\  
1995 Aug 31 & 49960.588 & 0.906 & 1.557 & 0.024 & 82.2 &  1.0\\  
1995 Sep  2 & 49962.833 & 0.683 & 1.684 & 0.020 & 80.4 &  1.0\\  
1995 Sep  3 & 49963.567 & 0.938 & 1.564 & 0.024 & 83.2 &  1.0\\  
1995 Sep  3 & 49963.774 & 0.010 & 1.616 & 0.024 & 82.1 &  1.0\\  
1995 Sep  4 & 49964.616 & 0.301 & 1.570 & 0.019 & 80.9 &  1.0\\  
1996 Jan 1 & 50084.490 & 0.824 & 1.775 & 0.031 & 84.4 &  1.0\\  
1996 Jan 3 & 50085.588 & 0.205 & 1.706 & 0.028 & 82.4 &  1.0\\  
1996 Jan 4 & 50087.483 & 0.861 & 1.643 & 0.028 & 81.9 &  1.0\\  
1996 Jan 7 & 50089.488 & 0.555 & 1.698 & 0.032 & 85.6 &  1.0\\  
1996 Jun 1 & 50235.768 & 0.225 & 1.630 & 0.024 & 81.8 &  1.0\\  
1996 Jun 2 & 50236.760 & 0.569 & 2.205 & 0.045 & 75.0 &  1.0\\  
1996 Jun 2 & 50236.787 & 0.578 & 1.590 & 0.031 & 81.0 &  1.0\\  
1996 Jun 12 & 50246.769 & 0.036 & 1.480 & 0.028 & 84.9 &  1.0\\  
1996 Jul 9 & 50273.682 & 0.358 & 2.108 & 0.041 & 71.4 &  1.0\\  
1996 Aug 24 & 50319.730 & 0.309 & 1.624 & 0.022 & 81.2 &  1.0\\  
1996 Aug 25 & 50320.592 & 0.607 & 1.549 & 0.025 & 82.0 &  1.0\\  
1996 Aug 25 & 50320.678 & 0.637 & 1.642 & 0.024 & 81.6 &  1.0\\  
1996 Aug 25 & 50320.808 & 0.682 & 1.721 & 0.028 & 81.2 &  1.0\\  
1996 Sep 1 & 50327.630 & 0.045 & 1.557 & 0.033 & 80.6 &  1.0\\  
1996 Sep 1 & 50327.742 & 0.084 & 1.699 & 0.029 & 80.3 &  1.0\\  
1996 Sep 1 & 50327.853 & 0.123 & 1.931 & 0.032 & 82.4 &  1.0\\  
1996 Sep 4 & 50330.614 & 0.079 & 1.626 & 0.027 & 81.2 &  1.0\\   
1996 Sep 4 & 50330.720 & 0.116 & 1.598 & 0.027 & 80.0 &  1.0\\  
1996 Sep 4 & 50330.846 & 0.159 & 1.644 & 0.029 & 80.8 &  1.0\\  
1996 Sep 6 & 50332.583 & 0.761 & 1.610 & 0.027 & 79.9 &  1.0\\  
1996 Sep 7 & 50333.563 & 0.100 & 1.565 & 0.028 & 83.8 &  1.0\\  
1996 Sep 7 & 50333.568 & 0.102 & 1.606 & 0.033 & 84.0 &  1.0\\  
1996 Sep 7 & 50333.685 & 0.143 & 1.536 & 0.027 & 84.0 &  1.0\\  
1996 Sep 7 & 50333.820 & 0.189 & 1.597 & 0.029 & 83.9 &  1.0\\  
1996 Sep 8 & 50334.587 & 0.455 & 1.354 & 0.030 & 84.8 &  1.0\\  
1996 Sep 8 & 50334.715 & 0.499 & 1.532 & 0.025 & 73.6 &  1.0\\  
1996 Sep 8 & 50334.721 & 0.501 & 1.575 & 0.029 & 83.5 &  1.0\\  
1996 Oct 12 & 50368.637 & 0.250 & 1.469 & 0.021 & 83.4 &  1.0\\  
1996 Oct 16 & 50372.532 & 0.599 & 1.597 & 0.022 & 81.8 &  1.0\\  
1996 Dec 22 & 50439.523 & 0.804 & 1.727 & 0.027 & 81.2 &  1.0\\  
1997 Jun 3  & 50602.791 & 0.358 & 1.374 & 0.047 & 81.5 &  1.0\\  
1997 Jun 4  & 50603.799 & 0.707 & 1.581 & 0.044 & 80.9 &  1.0\\  
1997 Jun 5  & 50604.772 & 0.044 & 1.514 & 0.033 & 81.5 &  1.0\\  
1997 Jun 6  & 50605.784 & 0.395 & 1.452 & 0.027 & 80.5 &  1.0\\  
1997 Jun 7  & 50606.786 & 0.742 & 1.465 & 0.035 & 79.0 &  1.0\\  
1997 Jun 9  & 50608.774 & 0.431 & 1.434 & 0.028 & 79.7 &  1.0\\  
1997 Jun 15 & 50614.773 & 0.509 & 1.460 & 0.030 & 79.7 &  1.0\\  
1997 Jun 16 & 50615.756 & 0.849 & 1.679 & 0.036 & 80.0 &  1.0\\  
1997 Jul 6 & 50635.753 & 0.776 & 1.525 & 0.025 & 79.3 &  1.0\\  
1997 Jul 8 & 50637.788 & 0.481 & 1.330 & 0.028 & 83.3 &  1.0\\  
1997 Jul 10 & 50639.777 & 0.170 & 1.693 & 0.037 & 82.2 &  1.0\\  
1997 Jul 11 & 50640.674 & 0.480 & 1.693 & 0.027 & 81.1 &  1.0\\  
1997 Aug 25 & 50685.723 & 0.085 & 1.605 & 0.017 & 83.9 &  1.0\\  
1997 Sep 6 & 50697.623 & 0.207 & 1.816 & 0.027 & 78.2 &  1.0\\  
1997 Sep 8 & 50699.823 & 0.969 & 1.536 & 0.033 & 81.6 &  1.0\\  
1997 Sep 9 & 50700.575 & 0.229 & 1.624 & 0.024 & 81.1 &  1.0\\  
1997 Sep 9 & 50700.707 & 0.275 & 1.673 & 0.024 & 80.5 &  1.0\\  
1997 Sep 9 & 50700.820 & 0.314 & 1.553 & 0.024 & 79.7 &  1.0\\  
1997 Oct 31 & 50752.532 & 0.227 & 1.630 & 0.030 & 81.4 &  1.0\\  
1997 Nov 4 & 50756.600 & 0.636 & 1.723 & 0.030 & 81.7 &  1.0\\  
1998 Jun 12 & 50976.666 & 0.864 & 1.505 & 0.108 & 78.0 & 2.1\\
1998 Jul 15 & 51009.749 & 0.324 & 1.864 & 0.056 & 77.1 & 1.0\\
1998 Aug 28 & 51053.765 & 0.571 & 1.662 & 0.029 & 81.9 &  1.0\\  
1999 Jun 11 & 51340.795 & 0.995 & 1.532 & 0.046 & 84.1 &  1.0\\  
1999 Jun 12 & 51341.770 & 0.332 & 1.734 & 0.046 & 81.4 &  1.0\\  
1999 Jun 13 & 51342.758 & 0.675 & 1.879 & 0.043 & 82.0 &  1.0 
\tablenotetext{1}{Calculated with the ephemeris $2445607.374
+2.886926E$ (Grankin et al. 1992).}
\enddata
\end{deluxetable}

\newpage
% ---- V DATA: litterature, Pic and Megantic --------------------------
\begin{deluxetable}{lcccccccc}
\tablewidth{0pt}
\tablecaption{Polarization data for MWC~1080, $V$ filter \label{Tab-Pic-V}}
\tablehead{
\colhead{UT Date} & \colhead{Aperture} & \colhead{JD} &
\colhead{Phase\tablenotemark{1}} &  
\colhead{P(\%)} & \colhead{$\sigma(P)$} &
\colhead{$\theta(\arcdeg)$} & \colhead{$\sigma(\theta)$} & Ref.\\
\colhead{} & \colhead{} & 2400000.0$+$ & \colhead{} &
\colhead{} & \colhead{} &
\colhead{} & \colhead{} & \colhead{}}
\startdata
Oct 73 & 10\arcsec & \nodata     & \nodata & 2.54 & 0.3 & 68.2 & \nodata & 1\\
Jan 76 & 13\arcsec &\nodata     & \nodata & 1.87 & 0.15 & 74 & \nodata & 2 \\ 
\nodata & 10\arcsec &\nodata & \nodata & 1.79 & \nodata & 75.0 & \nodata & 3\\
1996 Oct 9  & 10\arcsec &50366.464 & 0.497 & 1.882 & 0.038 & 77.0 & 1.0 & 4\\
1996 Oct 10 & 10\arcsec &50367.373 & 0.812 & 1.985 & 0.043 & 75.6 & 1.0 & 4\\
1996 Oct 11 & 10\arcsec &50368.402 & 0.168 & 2.110 & 0.062 & 78.5 & 1.0 & 4\\
1996 Oct 12 & 10\arcsec &50369.470 & 0.538 & 2.103 & 0.057 & 78.9 & 1.0 & 4\\
1996 Oct 16 & $8\farcs2$ &50372.589 & 0.619 & 2.044 & 0.049 & 75.1 & 1.0 & 5\\
1997 Jul 11 & $8\farcs2$ &50640.738 & 0.503 & 1.713 & 0.047 & 75.9 & 1.0 & 5\\
1998 Jul 3  & $8\farcs2$ &50997.649 & 0.133 & 2.119 & 0.068 & 74.0 & 1.0 & 5
\tablenotetext{1}{Calculated with the ephemeris $2445607.374
+2.886926E$ (Grankin et al. 1992).}
\enddata
\tablerefs{1. Vrba et al. 1975; 2. Garrison \& Anderson 1978;
3. Hillenbrand et al. 1992; 4. This paper. Data obtained at the
Observatoire du Pic-du-Midi. 5. This paper. Data obtained at the
Observatoire du Mont M\'egantic.} 
\end{deluxetable}

%\newpage
% ---------------------------- VARIABILITY ---------------------------
%\begin{deluxetable}{lccccccc}
%\tablewidth{0pt}
%\tablecaption{Results of the variability tests for MWC~1080. \label{Tab-Var}}
%\tablehead{
%\colhead{Wavelength} & \colhead{$N_{\rm obs}$} & \colhead{} & 
%\colhead{$\sigma_{\rm sample}$} & \colhead{$\sigma_{\rm mean}$} &
%\colhead{$Z \pm \sigma_Z$} & \colhead{$P{\chi^2}$} &
%\colhead{$P{\chi^2}$}\\
%\colhead{} & \colhead{} & \colhead{} & 
%\colhead{} & \colhead{} &
%\colhead{} & \colhead{$1\sigma$} &
%\colhead{$1.5\sigma$}}
%\startdata
%$7660\;$\AA &62&$Q$& 0.1033 & 0.0035 & 3.27 0.09 & 1.00 & 1.00\\
%        &  &$U$& 0.1027 & 0.0035 & 3.38 0.09 & 1.00 & 1.00\\
%$5500\;$\AA &7 &$Q$& 0.1509 & 0.0186 & 2.98 0.29 & 1.00 & 1.00\\
%        &  &$U$& 0.1269 & 0.0186 & 2.27 0.29 & 1.00 & 0.99
%\enddata
%\end{deluxetable}

\newpage
% ---------------------------- P versus lambda --------------------- 
\begin{deluxetable}{lcccccccc}
\tablewidth{0pt}
\tablecaption{Polarization as a function of the
wavelength \label{Tab-P_lambda}}
\tablehead{
\colhead{UT Date} & \colhead{JD} & \colhead{Phase\tablenotemark{1}} & 
\colhead{Filter} &
\colhead{P(\%)} & \colhead{$\sigma(P)$} &
\colhead{$\theta(\arcdeg)$} & \colhead{$\sigma(\theta)$}\\
\colhead{} & 2400000.0$+$ & \colhead{} &
\colhead{} & 
\colhead{} & \colhead{} &
\colhead{} & \colhead{}}
\startdata
1996 Oct 16 & 50372.626 & 0.632 & 4350(990)  & 1.910&0.225&70.9&3.4\\
             & 50372.589 & 0.619 & 5480(1060) & 2.044&0.049&75.1&1.0\\
             & 50372.532 & 0.600 & 7660(2410) & 1.597&0.022&81.8&1.0\\
             & 50372.561 & 0.609 & 8580(630)  & 1.441&0.050&75.4&1.0\\
1997 Jul 11 & 50640.711 & 0.494 & 4350(990)  & 1.632&0.116&72.4&2.0\\
             & 50640.738 & 0.503 & 5480(1060) & 1.713&0.047&75.9&1.0\\ 
             & 50640.787 & 0.520 & 5900(800)  & 1.786&0.053&79.1&1.0\\
             & 50640.808 & 0.527 & 7610(915)  & 1.439&0.053&83.4&1.0\\ 
             & 50640.762 & 0.511 & 8580(630)  & 1.400&0.049&85.6&1.0\\
1998 Jul 3 & 50997.712 & 0.155 & 4350(990) &  1.945&0.157&76.7&2.3\\
            & 50997.649 & 0.133 & 5480(1060)&  2.119&0.068&74.0&1.0\\
            & 50997.763 & 0.173 & 8580(630) &  1.533&0.072&81.5&1.3\\
            & 50997.793 & 0.183 & 7610(915) &  1.558&0.046&79.5&1.0\\
\tablenotetext{1}{Calculated with the ephemeris $2445607.374
+2.886926E$ (Grankin et al. 1992).}
\enddata
\end{deluxetable}

\end{document}